\documentclass[12pt,letter]{article}
\usepackage{graphicx,subfigure}
\usepackage{epsfig,amssymb}
mm\evensidemargin=-5true mm \topmargin=-15true mm

\newcommand{\beq}{\begin{equation}}
\newcommand{\eeq}{\end{equation}}
\newcommand{\be}{\begin{equation}}
\newcommand{\ee}{\end{equation}}
\newcommand{\beqa}{\begin{eqnarray}}
\newcommand{\eeqa}{\end{eqnarray}}
\newcommand{\beqar}{\begin{eqnarray*}}
\newcommand{\eeqar}{\end{eqnarray*}}
\newcommand{\bea}{\begin{eqnarray}}
\newcommand{\eea}{\end{eqnarray}}

\newcommand{\p}{\partial}
\newcommand{\al}{\alpha}

\newcommand{\n}{\nabla}
\newcommand{\s}{\sigma}

\newcommand{\ie}{{\it i.e.,}\ }

\begin{document}

\setlength{\unitlength}{1mm}

\begin{titlepage}

\begin{flushright}
arXiv: hep-th/0703251
\end{flushright}
\vspace{1cm}

\begin{center}
{\bf \Large Holographic Renormalization of Gravity in Little String Theory Duals}
\end{center}

\vspace{1cm}

\begin{center}
Donald Marolf and Amitabh Virmani

\vspace{0.5cm} {\small {\textit{Department of Physics, University of
California, Santa Barbara, CA 93106-9530, USA}}}

\vspace*{0.5cm} {\tt marolf@physics.ucsb.edu,
virmani@physics.ucsb.edu}
\end{center}

\vspace{1cm}

\begin{abstract}
We perform a holographic renormalization of gravity duals to little string theories.
In particular, we construct counterterms which yield a well-defined type II action for NS-sector  linear dilaton backgrounds.  Our methods are based on a similar recent construction in asymptotically flat spacetimes, and our work demonstrates in detail the parallels between asymptotically flat and linear dilaton boundary conditions.   The counterterms guarantee that (i) the on-shell action is finite and (ii)
asymptotically linear dilaton solutions are stationary points of the action under all
boundary condition preserving variations. We use the resulting action to compute a boundary stress tensor and the associated conserved charges.
\end{abstract}

\end{titlepage}

\vspace{0.5cm} \tableofcontents


\setcounter{equation}{0}
\section{Introduction}

The discovery of holographic gauge/gravity dualities has been a key development in modern string theory.  Such dualities relate quantum theories of gravity in high dimensions to quantum theories in lower dimensions on fixed metric backgrounds. The best understood such dualities occur in AdS/CFT \cite{Juan} and matrix theory \cite{BFSS}, in which the non-gravitating theories are local.    Such dualities provide new insights into both gravity and, in the case of AdS/CFT,  into strongly coupled gauge theories.

Another well known, but less well-studied class of dualities relates little string theories to asymptotically linear dilaton spacetimes \cite{ABKS,NSNS}.  Interestingly, little string theories are {\it non}-local \cite{NSNS} (see \cite{Ofer,Kutasov:2001uf} for reviews).  On general grounds, one might expect non-locality to be generic within the set of all holographic duals to gravitating bulk theories.  This already strongly motivates the study of holography in linear dilaton backgrounds, in an attempt to gain insights into the more generic situation\footnote{One might also study non-commuting theories with gravity duals\cite{HI,MR,CO,HRS}.}. A more specific motivation, however, is the similarity between linear dilaton asymptotics and asymptotic flatness.  For example, the conformal completion of a linear dilaton spacetime has a null boundary. This similarity has been noted in various contexts \cite{Shiraz,PP,unpub}, and was recently reemphasized in \cite{AF} where it was used to motivate a framework for gauge/gravity dualities in asymptotically flat spacetimes.

In this work, we reverse the flow of information, using techniques \cite{MM, MMV} developed for asymptotically flat spacetimes to holographically renormalize the action for certain linear dilaton spacetimes.  Our main interest is in ten dimensional spacetimes with  linear dilaton asymptotics dual to 5+1 little string theory with $SO(4)$ internal symmetry, associated with an $S^3$ factor in the bulk.  Through this $S^3$ pass $N$ units of flux of the NS-gauge field strength $H_3=dB_2$.  We develop a set of counterterms which guarantee that (i) the on-shell action is finite and (ii) asymptotically linear dilaton solutions are stationary points of the action under all boundary condition preserving variations.  We also study the boundary stress tensor and other `response' functions defined by this action and show that they lead to standard results \cite{Witten, Gibbons:1992rh, Nappi, Creighton:1995uj, Davis:2004xi, Grumiller:2007ju, silence} for the conserved charges of thermally excited linear dilaton spacetimes.

The plan of the paper is as follows.  Section \ref{prelims} briefly reviews the relevant linear dilaton backgrounds and introduces the framework for calculations in the remaining sections.  Section \ref{toy} performs a holographic renormalization for a toy model of the system of interest, obtained by removing the $S^3$ factor and replacing the NS gauge field with an exponential potential for the dilaton.  This toy model illustrates the main features of the full ten-dimensional system, but greatly simplifies the equations.  We then analyze the full ten-dimensional system in section \ref{NS}.   In particular, we construct counterterms for the NS sector of Type II supergravity with our asymptotics. We also construct the renormalized stress tensor and an analogous scalar response function and discuss the conserved charges. In section \ref{app:thermo}, as an application of our formalism, we discuss thermodynamics of finite temperature little string theory.  Finally we close with some discussion in section \ref{conclusions}.

\setcounter{equation}{0}

\section{Review of Linear Dilaton Backgrounds}
\label{prelims}
\label{LSTrev}

In this section we briefly review the connection between little string theories and linear dilaton backgrounds, and state a definition of linear dilaton asymptotics.  We also note the similarity between the Einstein-frame description of linear dilaton spacetimes and Minkowski space.  Due to this similarity, one expects that techniques developed for asymptotically flat gravity will prove useful in the study of linear dilaton spacetimes.  Indeed, in sections \ref{toy} and \ref{NS} below,  we follow the construction and analysis of counterterms described in \cite{MM, MMV} for the gravitational action in asymptotically flat spacetimes\footnote{These methods were in turn inspired by the construction of AdS gravitational counterterms, see e.g.   \cite{skenderis,kraus}.  See also \cite{Mann,KLS,eeA,ABL} for other asymptotically flat counterterms.}.

Many gauge/gravity dualities can be constructed by taking decoupling limits of various brane configurations \cite{Juan,IMSY,JPMLC}.  In the bulk, this limit typically yields the near-horizon limit of some BPS brane solution. In particular, it was shown in \cite{NSNS} that the decoupling limit on coincident NS5 branes yields a linear dilaton spacetime associated with the NS5 `throat.'   The non-gravitating dual is a low-energy limit of the NS5-brane theory which, in contrast to the more familiar D3-brane context, retains a full tower of excited open strings and, as a result, does not reduce to a local field theory.

Our focus here will be on the bulk gravitating description of the decoupling limit.
In the string frame, the near-horizon description of
$N$ coincident NS5-branes takes the familiar form
\bea
ds^2_{string} &=& dx_6^2 + N \al' \left( d \s^2 + d \Omega^2_3 \right), \\
\Phi &=& - \s, \\
H_3 &=& 2 N \al' \epsilon_{S^3},
\label{bg:string}
\eea
where $dx_6^2$ is the 6-dimensional Lorentzian metric and $\alpha'  = \ell^{-2}_s$ sets
the string tension. Here $\Phi$ is the ten-dimensional dilaton and $H_3= dB_2$, where $B_2$ is the NS-NS two-form potential.   In the strong coupling regime at large
negative $\sigma$, the physics is more properly described by the
near-horizon metric of $N$ M5-branes on an $S^1$. However, we will
be interested only in the asymptotics at large $\sigma$ where M-theoretic
corrections are heavily suppressed.

We wish to holographically renormalize the NS-sector of the bulk action, including in particular the gravitational terms.  As a result, it is natural to work in the Einstein frame.  In this frame (\ref{bg:string}) takes the tantalizing form \cite{AF}
\bea
ds^2 &=& d \rho^2 + \rho^2 \left( dy_6^2 + \frac{1}{16} d\Omega_3^2 \right), \label{bg1} \\
\Phi &=& - 4 \ln \frac{\rho}{4 \sqrt{N \al'}}, \label{bg2}\\
H_3 &=& 2 N \al' \epsilon_{S^3}. \label{bg3}
\eea
where $\rho = 4 e^{\frac{\s}{4}} \sqrt{N \al'}$ and $y^i = \frac{x^i }{4 \sqrt{N \al'}}$.   While
(\ref{bg3}) is not asymptotically flat, one may note that the metric components involve the same
powers of $\rho$ as for flat Minkowski space in hyperbolic coordinates:
 \begin{equation}
 \label{hyp}
 ds^2_{Mink} = d \rho^2 + \rho^2 \omega_{ij} d\eta^i d\eta^j,
 \end{equation}
where $\rho^2 = x_ax^a$, $\omega_{ij}$ is the metric on the unit
$(d-1)$-dimensional Lorentz-signature hyperboloid ${\cal H}$, and
$\eta^i$ are coordinates on ${\cal H}$.   It is the  similarity of (\ref{bg1}) and (\ref{hyp}) which leads one to expect that techniques developed for asymptotically flat gravity will prove useful in the study of linear dilaton spacetimes.

We wish to consider general classes of spacetimes which approach solutions similar to (\ref{bg1}-\ref{bg3}) at large values of some radial coordinate $\rho$.  For generality, let the coordinates $x^i$ parametrize the hypersurface $\Sigma_{\rho}$ at any fixed value of $\rho$, which we take to have dimension $d-1$. If $n^\mu$ is the unit (outward-pointing) normal to $\Sigma_{\rho}$, then the induced metric on  $\Sigma_{\rho}$ is
\be
h_{\mu \nu} = g_{\mu \nu}- n_{\mu} n_{\nu},
\ee
and the Einstein frame metric on the bulk spacetime can be decomposed as
\be
ds^2  = g_{\mu \nu} dx^\mu dx^\nu = \left( N^2 + N_{i} N^{i} \right) d\rho^2 + 2 N_i dx^i d\rho + h_{ij} dx^i dx^j \label{met}.
\ee
Here $N$ and $N^i$ are analogous to the `lapse' and `shift' functions of the ADM decomposition \cite{Wald}, and the metric (\ref{met}) describes the spacetime in a general gauge.  Since our spacetime should asymptotically resemble (\ref{bg1}), we impose $N \rightarrow 1, N^i \rightarrow 0$.

In analogy with established methods in asymptotically flat spacetime \cite{AshtekarHansen, BS, B, skenAF} we consider the case where each function in (\ref{met}) can be expanded in an asymptotic series in $1/\rho$, at least to an order to be specified below.   In fact, we will find in sections \ref{toy} and \ref{NS} below that for the cases of interest ($d > 4$), we may impose $N = 1 + {\cal O}(\rho^{-2}), N^i = {\cal O}(\rho^{-2})$.  As a result, one may find a diffeomorphism which sets $N=1 , N^i  = 0$ identically, so that $\rho$ is a Gaussian normal coordinate\footnote{For $d=4$ one has $N = 1 + {\cal O}(1/\rho)$ so that such a diffeomorphism would introduce logarithmic terms in the expansion of the metric $h_{ij}$ and spoil the assumption above.}.

Thus, our Einstein frame metric admits an expansion of the form
\bea ds^2 &=&  d\rho^2 + h_{ij} dx^i dx^j  \label{metric1}\\
&=& d\rho^2 + \rho^2 \left( h^0_{ij}
 + \frac{h^1_{ij}}{\rho^\ell} +  \frac{h^2_{ij}}{\rho^{\ell+1}} +\ldots \right),
 \label{falloff1}
 \eea
 where $\ell \ge 2$ is an integer to be determined separately for each particular model below.
We assume that the dilaton $\Phi$ and three form $H_{\mu \nu \s}$ can similarly
be expanded in the form \bea \Phi &=& -4 \ln \rho + 4 \ln 4 +
\frac{\al(x)}{\rho^\ell} + \frac{\al_2(x)}{\rho^{\ell+1}} +\ldots  \label{falloff2} \\
\beta_{ijk} &:=& H_{ijk} = 2 N \al' \epsilon_{S^3} + \frac{\beta^1_{ijk}}{\rho^\ell} +\frac{\beta^2_{ijk}}{\rho^{\ell+1}} + \ldots  \label{falloff3}\\
\alpha_{ij} &:=& H_{\perp ij} = \frac{\al^1_{ij}}{\rho^\ell} +\frac{\al^2_{ij}}{\rho^{\ell+1}} + \ldots  \label{falloff4},
\eea
where we have introduced the electric ($\alpha_{ij}$) and magnetic  ($\beta_{ijk}$)  parts of
$H_{\mu \nu \s}$. We will understand ``asymptotically linear dilaton spacetimes'' to be of the form Eq.~(\ref{metric1}-\ref{falloff4}) for values of $\ell$ described below.

\setcounter{equation}{0}
\section{A Toy Model}
\label{toy}

In this section we consider a model graviton-dilaton system whose solution set includes the linear dilaton spacetime, but without the complication of the three-form gauge field $H=dB$ which is present in the full type II system.   Our toy model is formulated on a spacetime which is topologically ${\mathbb R}^{10}$  and, in the Einstein frame, the (unrenormalized) action is
\be S = \int_M
\sqrt{-g} \left[ R - \frac{1}{2}(\n \Phi)^2 + 4
e^{\frac{\Phi}{2}} \right] + 2 \int_{\partial M} \sqrt{-h} K,  \label{action1}\ee
where $R$ is the spacetime Ricci scalar and $\Phi$ is the ten dimensional dilaton. The boundary term is the standard Gibbons-Hawking term where $K$ is the trace of the extrinsic curvature of the boundary at spatial infinity.
The field equations for this action are \bea R_{ab} & = & \frac{1}{2} \n_a \Phi \n_b \Phi - \frac{1}{2}  g_{ab} e^{\frac{\Phi}{2}}, \label{toy:eom1}\\
\Box \Phi  &=& - 2 e^{\Phi/2}. \label{toy:eom2}\eea The linear dilaton spacetime
\bea
ds^2_{string} &=& d \s^2 + d x_i dx^i, \label{x1}\\
\Phi &=& - \s
\eea
is a solution to this model, where $ds^2_{string} = e^{ \Phi/2} ds^2_{Einstein}$  is the metric in the string frame and the  $x^i$ denote 9 flat directions. In the Einstein frame, this solution takes the form
\bea
ds^2  &=& d \rho^2 + \rho^2 (d y_i dy^i) \label{linear1}\\
\Phi &=& - 4 \ln \rho + 4 \ln 4,
\label{linear2}
\eea
where $\rho = 4 e^{\frac{\s}{4}}$ and $y^i = x^i/4$.

We wish to construct counter-terms which, when added to the action (\ref{action1}) make it
 both finite and  stationary on all asymptotically linear dilaton solutions.  As stated in section \ref{LSTrev}, an asymptotic linear dilaton spacetime should be of the form  (\ref{falloff1},\ref{falloff2}) for some $\ell$.  However, we need to choose a physically appropriate value of $\ell$, following the usual guidelines that i) $\ell$ be small enough to allow a physically interesting class of solutions and ii) $\ell$ be large enough that the phase space is well-defined.

 The correct answer can also be read off from a physically interesting class of solutions.  In particular, $\mathbb{R}^8$ times Witten's two dimensional black hole is naturally taken to describe thermal excitations of the above vacuum.  In Einstein frame, this solution takes the form
\bea
ds^2 &=& e^{\frac{a}{4}} \sqrt{\cosh \s} \left[ - \tanh^2 \s dt^2 + d \s^2 + \sum_{i=1}^{8}dx_i dx^i \right], \label{2dBH1}\\
\Phi &=& - \ln \cosh \s - \frac{a}{2}. \label{2dBH2}
\eea
It can be readily verified that this solution asymptotes to (\ref{linear1}-\ref{linear2}) with falloff conditions given by (\ref{falloff1},\ref{falloff2}) with $\ell=8$, and one expects other interesting solutions to have the same fall-off properties.
We shall henceforth require asymptotically  linear dilaton solutions of our toy model to be of the form  (\ref{falloff1},\ref{falloff2}) for $\ell\ge 8$. We will see below that this guarantees that the conserved quantities are finite.

\subsection{Covariant Counterterms}
\label{toy:counterterms}

The onshell value of the action $(\ref{action1})$ diverges for the background (\ref{linear1}--\ref{linear2}) and, as will see below, (\ref{action1}) also fails to be fully stationary on all asymptotically linear dilaton solutions.  However, both of these problems may be cured by adding
appropriate covariant counterterms on the boundary, in the spirit of holographic renormalization \cite{KS3, KS2, KS1}.

Our counter-term construction uses techniques developed for asymptotically flat gravity in \cite{MM, MMV}. Here we outline some essential elements of the construction. The counter-terms proposed in \cite{MM} are based on the Gauss-Codazzi equations. In particular, the renormalized action of \cite{MM} takes the form
\begin{equation}
\label{AFEHaction} S =  \int_M \sqrt{-g}R +
2 \int_{\partial M} \sqrt{-h} (K - \hat K),
\end{equation}
where
$K_{ij} = h_i{}^{k} \n_k n_j$ is the extrinsic curvature of the boundary $\partial M$ (considered as a surface of large constant $\rho$) which gives the familiar Gibbons-Hawking boundary term, and the ``counterterm'' is constructed from
$\hat K := h^{ij} \hat K_{ij}$. This $\hat K_{ij}$
may be roughly thought of as `the extrinsic curvature which would be obtained if the boundary were embedded into flat Minkowski space.'  More precisely, $\hat K_{ij}$ is defined
to satisfy
\begin{equation}
\label{Khat}  {}{\cal R}_{ij} = \hat K_{ij} \hat K - \hat K_i^m \hat
K_{mj},
\end{equation}
where ${}{\cal R}_{ij}$ is the Ricci tensor of the boundary metric
$h_{ij}$ on $\partial M$ and we follow the conventions of
Wald \cite{Wald}.  The point is that (\ref{Khat}) has the same form as a Gauss-Codazzi equation for an embedding of the boundary into a spacetime of vanishing Riemann curvature.  Thus, when the boundary can be embedded into a flat spacetime, the extrinsic curvature of this embedding is indeed $\hat K_{ij}$.  However,  even when such an embedding is not possible (the generic situation in 4 or more spacetime dimensions), equation (\ref{Khat}) can be solved to yield a useful counter-term for the action.

Our linear dilaton counterterms take a similar form:
\be
S^{ct} = S^{ct}_{\hat K} + S^{ct}_{\Phi} = - 2 \int_{\p M} \sqrt{-h}  \hat K + \frac{1}{2} \int_{\p M} \sqrt{-h}  e^{\frac{\Phi}{4}}.
\label{actioncounter1}
\ee
Here $\hat K := h^{ij} K_{ij}$ and $\hat K_{ij}$ is now re-defined to satisfy
\be
{}{\cal R}_{ij} + \frac{1}{2} e^{\frac{\Phi}{2}} h_{ij} = \hat K_{ij} \hat K - \hat K_{ik}\hat K^k{}_j,
\label{hatK:defn1}
\ee
where ${}{\cal R}_{ij}$ is the Ricci tensor of the boundary metric $h_{ij}$. The extra term $\frac{1}{2} e^{\frac{\Phi}{2}} h_{ij} $ in (\ref{hatK:defn1}) guarantees that $\hat K_{ij}$ and $K_{ij}$ agree to leading order.
Below, we show in section \ref{toyonshell} that the resulting total action is finite on shell, and in section \ref{toyvary} that it is stationary on solutions when varied within the class of asymptotically linear dilaton spacetimes.

\subsubsection{The On-shell Action}
\label{toyonshell}

We now address the divergences of the on-shell action. The Gibbons-Hawking term and the bulk terms both diverge, and the divergences do not cancel.   Considering first the Gibbons-Hawking term, one notes that expanding (\ref{hatK:defn1}) in powers of $\rho$ yields
\be
\hat K_{ij} = K_{ij} (1 + {\cal O}(\rho^{-8})) = \rho h^0_{ij} (1 + {\cal O}(\rho^{-8})).
\ee
But the leading divergence in the Gibbons-Hawking term is only of order $\rho^8$.  Thus, all on-shell divergences from the Gibbons-Hawking boundary term are exactly cancelled by the $\hat K$ counterterm for any asymptotically linear dilaton solution.

Similarly, for asymptotically linear dilaton solutions the on-shell bulk action is
\be
S_{onshell}^{bulk} = - \int_{M} \sqrt{-g} e^{\frac{\Phi}{2}} = -2 \rho^8 V_{9} + 8 (\ln \rho) \int_{\p M} \sqrt{h^0} (h^1 + \al) + \mbox{ finite}. \label{onshell}
\ee
Now, as described in appendix \ref{sec:eom1order}, the perturbative equations of motion imply the relation
\be
{\cal R}^1_{ij} = - 4 ( \al+ h^1) h^0_{ij}.
\ee
Using this result together with the relation (\ref{waldeq}) one sees that the logarithmically divergent term reduces to a total divergence on $\partial M$, and hence does not contribute.

The $\rho^8$ divergence in (\ref{onshell}) is precisely canceled by the second counterterm $S^{ct}_{\Phi}$,
\be
S^{ct}_{\Phi} = \frac{1}{2} \int_{\p M}  \sqrt{-h} e^{\frac{\Phi}{4}} = + 2 \rho^8 V_{9} + \mbox{ finite},
\ee
so that the total action is finite onshell.

\subsubsection{Variation of the Action}
\label{toyvary}

The total action obtained by adding the covariant counterterms to the toy model action is
\be
S = S_0 + S^{ct}_{\hat K} + S^{ct}_{\Phi},
\label{actiontotal}
\ee
where
\bea
S_0 &=& \int_M
\sqrt{-g} \left[ R - \frac{(\nabla \Phi)^2}{2} + 4
e^{\Phi/2} \right] + 2 \int_{\partial M} \sqrt{-h} K, \\
S_{\hat K}^{ct} &=& - 2 \int_{\partial M} \sqrt{-h} \hat K,  \\
S_{\Phi}^{ct} &=& \frac{1}{2} \int_{\partial M} \sqrt{-h} e^{\Phi/4} .
\eea
Having shown that the onshell action is finite, we now argue that the variations of this action vanish when our boundary conditions (\ref{falloff1}--\ref{falloff2}) are preserved.   Now, when the equations of motion hold, it is clear that the variation of the action can be written as a boundary term. However,  the point here is to show that the resulting boundary term vanishes as well.  By direct calculation one finds that the boundary terms are
\bea \delta S_0 &=& \int_{\partial M} \sqrt{-h} \pi_{ij}\delta
h^{ij}  -
\int_{\partial M } \sqrt{-h} (n^\mu\partial_\mu \Phi)\delta \Phi, \\
\delta S_{\hat K}^{ct} &=& -2 \int_{\partial M} \delta(\sqrt{-h} \hat K) , \\
\delta  S_{\Phi}^{ct} &=&  \frac{1}{8}\int_{\partial M}  \sqrt{-h} e^{\Phi/4} \delta \Phi -
\frac{1}{4}\int_{\partial M} \sqrt{-h} e^{\frac{\Phi}{4}} h_{ij}\delta
h^{ij} ,
\eea
where
\be
\pi_{ij} = K_{ij} - K h_{ij},
\ee
and
\be \delta(\sqrt{-h} \hat K) = \sqrt{-h}
\left(\frac{1}{2} \hat K h^{ij} \delta h_{ij} + \hat K_{ij} \delta
h^{ij} + h^{ij} \delta \hat K_{ij} \right). \ee

As in \cite{MM}, one may solve for $h^{ij} \delta \hat K_{ij}$ as a series in $\rho^{-1}$ by considering the variation of the defining equation (\ref{hatK:defn1}) and expanding in powers of $\rho^{-1}$.
One finds
\bea
\label{temp}
\delta \hat K_{ij} h^{ij} = -
\rho h^0_{ij} \delta h^{ij} + \mathcal{O}\left(\frac{\delta
h}{\rho^{\ell-1}} \right) + \frac{9}{4 \rho} \delta \Phi  +
\mathcal{O}\left(\frac{\delta \Phi}{\rho^{\ell+1}} \right) +
\frac{1}{16 \rho} h^{0ij} \delta {\cal R}_{ij} +
\mathcal{O}\left(\frac{\delta {\cal R}}{\rho^{\ell+1}} \right).
\eea
Substituting this into the variation of the $\hat K$ counterterm action and taking the limit  $\rho \to \infty$ yields
\be \delta
S^{ct}_{\hat K} =  9\rho^{10} \int_{\p M} \sqrt{-h^0} h^0_{ij}  \delta h^{ij} - \frac{9 \rho^8}{2} \int_{\p M} \sqrt{-h^0} \delta \Phi, \ee
where we have discarded the term
\be
\frac{1}{8 \rho} \int_{\p M} \sqrt{-h} h^{0ij} \delta {\cal R}_{ij},
\label{totalderivative1}
\ee
which is a total divergence on $\partial M$.
This last  fact may be seen by using the relation (see e.g. \cite{Wald})
\be
\delta {\cal R}_{ij} = - \frac{1}{2} h^{kl} D_i D_j \delta h^1_{kl} - \frac{1}{2}
h^{kl} D_k D_l \delta h^1_{ij} + h^{kl} D_k D_{(i} \delta h^1_{j)l}, \label{waldeq}\ee
where $D_i$ is the covariant derivatives with respect to the metric $h_{ij}$. It is clear that replacing $h^{0ij}$ in (\ref{totalderivative1}) by $h^{ij}$ would yield a total divergence, but $h^{0ij}$ and $h^{ij}$ differ by a term of sufficiently high order so as not to change (\ref{totalderivative1}) in the limit $\rho \rightarrow \infty$.

The other terms give
\bea
\delta S_{0} &=& - 8 \rho^{10}  \int_{\p M} \sqrt{-h^0} h^0_{ij}  \delta h^{ij} + 4 \rho^8 \int_{\p M} \sqrt{-h^0} \delta \Phi\\
\delta S_{\Phi}^{ct} &=& \frac{\rho^8}{2}  \int_{\p M} \sqrt{-h^0} \delta \Phi -  \rho^{10}  \int_{\p M} \sqrt{-h^0} h^0_{ij}  \delta h^{ij}.
\eea
Thus, we have
\bea
\delta S = \delta S_{0} + \delta S^{ct}_{\hat K} + \delta S^{ct}_{\Phi} = 0.
\eea
With our choice of counterterms, variations of the total action within the space of asymptotically linear dilaton spacetimes vanish on solutions of the field equations.

\subsection{The Boundary Stress Tensor}
\label{toy:renormalized}

The counterterms constructed in section \ref{toy:counterterms} are local functionals of the boundary fields $\{ h_{ij}, \Phi \}$ on $\p M$. In analogy with the AdS construction, one might like to obtain holographic one-point functions (such as the boundary stress tensor) by varying the on-shell action with respect to $\{ h_{ij}, \Phi \}$.  Such an approach should be possible, though it is complicated by the ``momentum-dependent renormalization'' (see e.g. \cite{PP}) characteristic of linear dilaton backgrounds (and of asymptotically flat space)\footnote{Since variations of the total action about solutions vanish when the variations preserve the asymptotically linear dilaton boundary conditions, this method also requires an extension of $S$ to other boundary conditions (at least perturbatively).}.

However, another approach \cite{BY} is to define a boundary stress tensor via the $\rho_0 \rightarrow \infty$ limit of variations of the cut-off actions $S_{\rho_0}$, defined by computing $S$ (with all counter-terms) for a spacetime $M$ given only by the region $\rho \le \rho_0$.    Here we follow the treatment of \cite{MM}. Specifically, we define
\be
\label{toyT}
T_{ij} (\rho)  = \frac{-2}{\sqrt{-h}} \frac{\delta S_{\rho_0}}{\delta h^{ij}}, \\
\ee
This stress tensor is conserved in the sense that
\be
\label{toycons}
D^i T_{ij} =0
\ee
on $\Sigma_\rho$ at each value of $\rho$. We also define an analogous scalar response function:
\be
\label{toyjPhi}
j_\Phi (\rho)  =  \frac{1}{\sqrt{-h}} \frac{\delta S_{\rho_0}}{\delta \Phi}.
\ee
This response function is related (up to appropriate momentum-dependent renormalization) to the one-point function for the operator ${\cal O}_{\Phi}$ dual to $\Phi$ in the little string theory.
Tracing through the analysis of variations in section \ref{toyvary} shows that (\ref{toyT}) and (\ref{toyjPhi}) are respectively of order $\rho^{-7}$ and $\rho^{-9}$  for asymptotically linear dilaton solutions.  Thus, it is natural to introduce
\be
\tilde T_{ij} = \lim_{\rho \rightarrow \infty} \rho^7 T_{ij}, \ \ \ {\rm and} \hspace{0.5cm}   \tilde j_\Phi = \lim_{\rho \rightarrow \infty} \rho^9 j_\Phi.
\ee
Taking the large $\rho$ limit of (\ref{toycons}), one notes that $D^0_{i} \tilde T^{ij} =0$, where $D^0_i$ is the torsion-free covariant derivative compatible with $h^0_{ij}$. After some calculation one finds (see Appendix \ref{onepoint:toy})
\bea
\tilde T_{ij} &=& \frac{1}{4} \left[ - (h^1 - 8 p^1 + 7 \hat p^1 + 2 \al)h^0_{ij} + h^1_{ij} - 8 p^1_{ij} + 7 \hat p^1_{ij} \right], \label{toyT1}\\
\tilde j_\Phi &=& \frac{1}{8} \left( 47 \al - 4 h^1 +  4 \hat p^1\right), \label{toyjPhi1}
\eea
where we have used the expansions (see also appendix B of \cite{MMV})
\bea
p_{ij} = \frac{1}{\rho} K_{ij} = h^0_{ij} + \frac{1}{\rho^\ell} p^1_{ij} +\frac{1}{\rho^{\ell+1}} p^2_{ij} + \ldots  \ \ {\rm and} \\
\hat p_{ij} = \frac{1}{\rho} \hat K_{ij} = h^0_{ij} + \frac{1}{\rho^\ell} \hat p^1_{ij} +\frac{1}{\rho^{\ell+1}} \hat p^2_{ij} + \ldots \ \ .
\eea

By the general arguments given in \cite{MM} (and based on those of \cite{HIM2}), these expressions lead to conserved quantities which generate the expected asymptotic symmetries.  In particular, the generator of an asymptotic translation $\xi^i$ is
\bea
Q[\xi] &:=&   \lim_{\rho \rightarrow \infty} \int_{C_\rho} \sqrt{-h_C} T_{ij} \xi^i n_{C_\rho}^j \label{Q1}, \\
&=&   \int_C   \sqrt{-h_{C}^{0}} \tilde T_{ij}  \xi^i n_C^j  \label{Q2},
\eea
where  $C_\rho$ form a family of Cauchy surfaces within the constant $\rho$ hypersurfaces $\Sigma_\rho$, such that $C = \lim_{\rho \rightarrow \infty} C_\rho$ is a Cauchy surface in the boundary $\partial M$.  In (\ref{Q1},\ref{Q2}) $n^i_{C_\rho}, n^i_C$ are unit normals to $C_\rho, C$ respectively in $(\Sigma_\rho,h_{ij})$ and  $(\partial M,h^0_{ij})$.   In section \ref{app:thermo} we verify that (\ref{Q2}) reproduces the mass of the thermally excited solution, and that $\tilde T_{ij}$ reproduces the pressure.


\setcounter{equation}{0}
\section{NS Sector of Type II Supergravity}
\label{NS}

We now turn to the asymptotically linear dilaton spacetimes that arise in the type II string theories dual to little string theories.  We work in the semi-classical bulk approximation, where the (unrenormalized) action with Gibbons-Hawking term takes the form
\bea S_0 = \int_{M} \sqrt{-g}  \left[ R - \frac{1}{2} \n_{\mu} \Phi
\n^{\mu} \Phi- \frac{1}{12} e^{-\Phi}
H_{\mu \nu \rho} H^{\mu \nu \rho}  \right] + 2 \int_{\p M} \sqrt{-h} K,
\label{action}
\eea
in the Einstein frame, where $\Phi$ is the dilaton and $H_3 = d B_2$ in terms of the  NS-NS two-form potential $B_2$.   For simplicity we have set the overall normalization ($16\pi G_N$) of the action (\ref{action}) to unity.  The resulting equations of motion are:
\bea
R_{\mu \nu} &=& \frac{1}{2} \n_{\mu} \Phi \n_{\nu} \Phi - \frac{1}{48} e^{-\Phi} H_{\s \kappa \gamma } H^{\s \kappa \gamma } g_{\mu \nu} + \frac{1}{4} H_{\mu \s \kappa } H_{\nu}{}^{\s \kappa}, \label{eom1} \\
\Box \Phi &=& - \frac{1}{12} e^{-\Phi} H_{\s \kappa \gamma } H^{\s \kappa \gamma }, \label{eom2} \\
\n_{\mu} \left( e^{-\Phi} H^{\mu \s \kappa} \right) &=& 0. \label{eom3}
\eea

The linear dilaton solutions were already discussed in section \ref{prelims}.  However, as in the case of our toy model, we need to fix the fall-off parameter $\ell$ in our definition of `asymptotically linear dilaton spacetimes.' Again, we will read off this variable $\ell$ from a physically interesting class of solutions describing thermal excitation over the background (\ref{bg1}-\ref{bg3}). Such solutions are given by (see for example \cite{MS}):
\bea
ds^2 &=& e^{\frac{a}{4}} \sqrt{\cosh \s} \left[- \tanh^2 \s dt^2 + N \al' \left( d\s^2 + d \Omega_3^2\right) + dx_5^2 \right]\label{thermal1}\\
 \Phi &=& - \ln \cosh \s - \frac{a}{2} \label{thermal2}\\
H_3 &=& 2 N \al' \epsilon_{S^3} \label{thermal3},
\eea
from which it follows that $\ell = 8$.

As in our toy model, counterterms must be added to $S_0$ to obtain an action which is both finite on-shell and stationary under all asymptotically linear dilaton variations.  The on-shell divergences are discussed below in section \ref{NS:counter} and used to motivate a particular choice of counter-terms which make the action finite on-shell.  We then show that the resulting action is stationary under all asymptotically linear dilaton variations in section \ref{NS:vary}.


\subsection{Covariant Counterterms and the On-shell Action}
\label{NS:counter}

Let us first consider the divergences in the bulk onshell action.  Using equation (\ref{eom1}) we find the bulk onshell action to be
\be
S_{bulk}^{onshell} = - \frac{1}{24} \int_{M} \sqrt{-g} e^{-\Phi} H^2,
\ee
which for the linear dilaton background (\ref{bg1}-\ref{bg3}) diverges as
\be
- \frac{2 V_9}{\epsilon^8}\label{bulk},
\ee
where $V_9$ is the volume of the boundary manifold ($V_9 = \frac{1}{4^3} V_6 (2 \pi^2)$) and $\rho = \frac{1}{\epsilon}$ is an IR cutoff. As in the case of our toy model, divergences from the Gibbons-Hawking boundary term do not cancel divergences from the bulk action. We need to add appropriate counterterms.

Motivated by our toy model, the first counterterm we add is a $\hat K$ counterterm
\be
S_{ct}^1 = -  2 \int_{\p M} \sqrt{-h} \hat K \label{hatK;defn2}
\ee
where this time $\hat K $ is the solution of the  following equation:
\be
\hat {\cal R}_{ij} := {\cal R}_{ij}  +  \frac{1}{2 (N \al')} e^{\Phi/2} h_{ij}  - \frac{1}{4}e^{-\Phi} \beta_{i}{}^{lm} \beta_{jlm} = \hat K_{ij} \hat K - \hat K_i^m \hat K_{mj}. \label{defn}
\label{rhat}
\ee
The first two terms in $\hat {\cal R}_{ij}$ match (\ref{hatK:defn1}), and the addition of the third term guarantees that $\hat K$ agrees with $K$ to leading order. In particular,
\be
\hat K_{ij} = K_{ij} = \rho h^0_{ij}  + {\cal O} \left( \frac{h^1_{ij}}{\rho^{\ell-1}}\right),
\ee
so that the onshell divergences from the Gibbons-Hawking boundary terms are exactly canceled by (\ref{hatK;defn2}).

We also add three additional  counterterms:
\bea
S_{ct}^2 &=& - \frac{(N\al')^{1/2}}{4} \int_{\p M} \sqrt{-h} e^{- \frac{5}{4} \Phi} \beta_{ijk} \beta^{ijk}, \\
S_{ct}^3 &=& + \frac{(N\al')^{3/2}}{16} \int_{\p M} \sqrt{-h} e^{- \frac{7}{4} \Phi} \beta_{imn} \beta_j{}^{mn} {\cal R}^{ij},\\
S_{ct}^4 &=&  + \frac{7}{2} \int_{\p M} \sqrt{-h} e^{ \frac{1}{4} \Phi}.
\eea
The divergent contributions are
\bea
S_{ct}^{2} &\sim& - \frac{24}{\epsilon^8} V_9, \label{ctcon1}\\
S_{ct}^{3} &\sim& \frac{12}{\epsilon^8} V_9, \label{ctcon3}\\
S_{ct}^{4} &\sim& \frac{14}{\epsilon^8} V_9 \label{ctcon3}.
\eea
Adding the bulk (\ref{bulk}) and the counterterm contributions (\ref{ctcon1}-\ref{ctcon3}) yields a finite onshell action.

\subsection{Variation of the Action}

\label{NS:vary}

Having shown that the total action
\bea
S_{tot} = S_0 + S_{ct}^1 + S_{ct}^2 + S_{ct}^3 + S_{ct}^4
\eea
is finite onshell, we now show that the variations of this action vanish when our boundary conditions (\ref{falloff1}- \ref{falloff4}) are preserved. Again when the equations of motion hold, it is clear that the variation of the action can be written as a boundary term. However, as was in the case of our toy model, the point is to show that the resulting boundary terms vanish as well. By direct calculation one finds that the variation of (\ref{action}) (after imposing the equations of motion) is
\bea
\delta S_0 =  \int_{\p M} \sqrt{-h}\pi_{ij} \delta h^{ij} - \int_{\p M} \sqrt{-h} (n^\mu \n_{\mu} \Phi) \delta \Phi - \frac{1}{6} \int_{\p M} \sqrt{-h} (n_\mu  H^{\mu ij}) \delta B_{ij} e^{-\Phi},
\eea
where $n_{\mu}$ is the unit normal pointing in the radial direction and \be \pi_{ij} = K_{ij} - h_{ij} K. \ee  For the background solution (\ref{bg1}-\ref{bg3}) one sees that $n_{\mu} H^{\mu \nu \s}$ is zero, so the last term in the variation does not contribute. To leading order the other two terms simplify to become
\be
\delta S_0 =  -  8 \rho^{10} \int_{\p M} \sqrt{-h^0} h_{ij}^0 \delta h^{ij} + 4 \rho^8 \int_{\p M}  \sqrt{-h^0} \delta \Phi.
\label{vari}
\ee
We immediately note that these terms are non-zero for field variations which preserve our boundary conditions. However, we will shortly see that the counterterms variations cancel these terms. The variation of the first counterterm $S_{ct}^1$ is
\bea
\delta S_{ct}^1 = -2 \int_{\p M} \delta \left( \sqrt{-h} \hat K\right) d^9 x
\eea
where
\bea
\delta \left( \sqrt{-h} \hat K \right) = \sqrt{-h} \left( \frac{1}{2} \hat K h^{ij} \delta h_{ij} + \hat K_{ij} \delta h^{ij} + h^{ij} \delta \hat K_{ij}\right).
\eea
As in the case of our toy model, one may solve for $h^{ij} \delta \hat K_{ij}$ as a power series in $\rho^{-1}$ by considering the variation of the defining equation (\ref{rhat}) and expanding in powers of $\rho^{-1}$. One finds,
\be
\delta \hat K_{kl} h^{kl} = - \frac{1}{2} \rho h^0_{kl} \delta h^{kl} + \frac{1}{16 \rho} h^{0ij} \delta \hat {\cal R}_{ij} + {\cal O} \left( \frac{h^1}{\rho^{\ell-1}}\right),
\ee
where the variation  $ \delta \hat {\cal R}_{ij}$ is given by
\bea
\delta \hat {\cal R}_{ij} &=& \delta {\cal R}_{ij} + \frac{1}{4 (N \al')} e^{\Phi/2} \delta \Phi h_{ij} + \frac{1}{2(N \al')} e^{\Phi/2} \delta h_{ij}\nonumber \\ &+& \frac{1}{4} e^{-\Phi} \beta_{i}{}^{lm} \beta_{jlm} \delta \Phi - \frac{1}{4} e^{-\Phi} \left(  \delta \beta_{ilm} \beta_{j}{}^{lm} + \beta_{i}{}^{lm} \delta \beta_{jlm} + 2 \beta_{ipn} \beta_{jm}{}^{n} \delta h^{pm}\right).
\eea
Thus,
\be
\delta S_{ct}^1 =  8 \rho^{10} \int_{\p M} \sqrt{-h^0} h_{ij}^0 \delta h^{ij}  - \frac{1}{8 \rho} \int_{\p M} \sqrt{-h} h^{0ij} \delta \hat {\cal R}_{ij} + \int_{\p M} \sqrt{-h} \left[{\cal O} \left( \frac{h^1}{\rho^{\ell +1}}\right)\right].
\ee
The first term exactly cancels the first term in the variation (\ref{vari}). The second term has various pieces and one of them is
\be
\frac{1}{8 \rho} \int_{\p M}  \sqrt{-h} h^{0ij} \delta {\cal R}_{ij},
\ee
where $\delta {\cal R}_{ij}$ is the variation of the boundary Ricci tensor.
To the requisite order the integrand is a total divergence (see section \ref{toyvary}), and gives no contribution.

We may further simplify $\delta S_{ct}$ by using the explicit form (\ref{bg1}-\ref{bg3}) of the asymptotic fields. To the leading order we find
\bea
\delta S +   \delta S_{ct}^1 &=& - \frac{25 \rho^8}{2 } \int_{\p M} \sqrt{-h^0} \delta \Phi + \rho^{10} \int_{\p M} \sqrt{-h^0} h^0_{ij} \delta h^{ij} + \frac{\rho^8}{4^6 (N\al')^2 } \int_{\p M } \sqrt{-h^0} \beta^{0\:imn} \delta \beta_{imn} \cr &+& \frac{\rho^{10}}{2} \int_{\p M} \sqrt{-h^0} \Omega^0_{mp} \delta h^{mp},
\label{term1}
\eea
where $\Omega^0_{ij}$ is the unit metric of $S^3$.
Similar calculations for the variations $\delta S_{ct}^2$ and $\delta S_{ct}^3$ give the result
\bea
\delta S_{ct}^2 + \delta S_{ct}^3 &=& 9\rho^8 \int_{\p M} \sqrt{-h^0} \delta \Phi - \frac{\rho^8}{4^6 (N\al')^2} \int_{\p M} \sqrt{-h^0}  \beta^{0\:ijk} \delta \beta_{ijk} + 6 \rho^{10} \int_{\p M} \sqrt{-h^0} h^0_{ij} \delta h^{ij} \cr & -& \frac{\rho^{10}}{2} \int_{\p M} \sqrt{-h^0} \Omega^0_{mp} \delta h^{mp}.
\label{term2}
\eea
Adding (\ref{term1}) and (\ref{term2}) yields
\be
\delta S +   \delta S_{ct}^1 +\delta S_{ct}^2 + \delta S_{ct}^3 = - \frac{7\rho^8}{2 } \int_{\p M} \sqrt{-h^0} \delta \Phi + 7 \rho^{10} \int_{\p M} \sqrt{-h^0} h^0_{ij} \delta h^{ij},
\ee
which cancel with the variation of $S_{ct}^4$. Thus, when the dust settles one finds
\be
\delta S_{tot} = 0.
\ee

\subsection{The Boundary Stress Tensor}
\label{NS:renormalized}

The counterterms constructed in section \ref{NS:counter} are local functionals of the boundary fields $h_{ij}, \Phi$, and $\beta_{ijk}$ on $\partial M$. In analogy with the AdS construction, one might like to obtain holographic one-point functions by varying the action with respect to these fields.  Such an approach should be possible, though it is complicated by the ``momentum dependent renormalization.''

Here we follow the treatment of section \ref{toy:renormalized}. Specifically, we use the  definitions (\ref{toyT}) and (\ref{toyjPhi}) for the boundary stress tensor and the scalar response function respectively. Again we introduce
\be
\tilde T_{ij} = \lim_{\rho \rightarrow \infty} \rho^7 T_{ij}, \ \ \ {\rm and}  \hspace{0.5cm}  \tilde j_\Phi = \lim_{\rho \rightarrow \infty} \rho^9 j_\Phi.
\ee
After some calculation one finds (see Appendix \ref{onepoint:NS})
\be
\tilde j_{\Phi}= \frac{1}{56} (-124 h^1+39 h^1 \cdot \Omega^0 +124 \hat p^1 -4 \left(\hat p^1 \cdot \Omega^0 + \frac{49}{256} {\cal R}^1 \cdot  \Omega^0 \right) +1001 \alpha
   )-\frac{\beta^{1 ijk}\beta^0_{ijk}}{16384}
   \label{jtildePhi:NS}
\ee
where $h^1\cdot \Omega^0$ is the contraction of $h^{1ij}$ with the unit metric $\Omega^0_{ij}$ on $S^3$  \ie
\be h^1\cdot \Omega^0 = h^{1ij} \Omega^0_{ij}, \ee
and similarly for $\hat p^1 \cdot \Omega^0$, etc. The corresponding expressions for $\tilde T_{ij}$ are exceedingly long. Instead of presenting an explicit expression, we simply note that $\tilde T_{ij}$ is the coefficient of $\frac{1}{\rho^7}$ in the expansion
\bea
T_{ij} &=& -2 \hspace{0.1cm} \bigg{(} \pi_{ij} + \hat K h_{ij} - 2 \hat K_{ij} - 2 M_{ij} - \frac{3}{4} e^{-\frac{5}{4} \Phi} (N \al')^{1/2} \beta_{imn} \beta_{j}{}^{mn} \nonumber \\ &+& \frac{1}{8}e^{-\frac{5}{4} \Phi} (N \al')^{1/2} \beta_{lmn} \beta^{lmn} h_{ij} - \frac{1}{32}e^{-\frac{7}{4}\Phi} \beta_{pmn}\beta_q{}^{mn} {\cal R}^{pq} (N\al')^{3/2} h_{ij} \nonumber \\ &+& \frac{1}{16} e^{-\frac{7}{4} \Phi} (N \al')^{3/2} \left[ 2 {\cal R}^{mn} \beta_{nip} \beta_{mj}{}^{p} + 2 {\cal R}_{(i|p|} \beta_{j)mn} \beta^{pmn} \right] - \frac{7}{4} e^{\frac{\Phi}{4}}h_{ij} + D_{ij} \bigg{)},
\label{stress}\eea
where  $M_{ij}$ and $D_{ij}$ are defined as
\bea
M_{ij} &:=& h^{kl} \frac{\delta}{\delta h^{ij}}\hat K_{kl} ,\\
D_{ij} &:=& \frac{(N\al')^{3/2}}{16}e^{-\frac{7}{4} \Phi}   \beta^{p}{}_{mn} \beta^{qmn} \frac{\delta}{\delta h^{ij}} R_{pq}.
\eea

By the general arguments given in \cite{MM} (and based on those of \cite{HIM2}), (\ref{stress}) lead to conserved quantities which generate the expected asymptotic symmetries.  In particular, the generator of an asymptotic translation $\xi^i$ is
\bea
Q[\xi] &:=&  \lim_{\rho \rightarrow \infty} \int_{C_\rho} \sqrt{-h_C} T_{ij} \xi^i n_{C_\rho}^j \label{QNS1}, \\
&=&  \int_C   \sqrt{-h_{C}^{0}} \tilde T_{ij}  \xi^i n_C^j  \label{QNS2},
\eea
where  $C_\rho$ form a family of Cauchy surfaces within the constant $\rho$ hypersurfaces $\Sigma_\rho$, such that $C = \lim_{\rho \rightarrow \infty} C_\rho$ is a Cauchy surface in the boundary $\partial M$.  In (\ref{QNS1},\ref{QNS2}) $n^i_{C_\rho}, n^i_C$ are unit normals to $C_\rho, C$ respectively in $(\Sigma_\rho,h_{ij})$ and  $(\partial M,h^0_{ij})$.
In section \ref{app:thermo} we verify that $T_{ij}$ gives the correct energy density and pressure for thermally excited linear dilaton solutions.

\setcounter{equation}{0}
\section{Thermodynamics}
\label{app:thermo}

As an application of our formalism, we now calculate the boundary stress tensor and conserved charges of the thermally excited linear dilaton spacetimes.  In particular, we will demonstrate that our formalism agrees with the standard results for the two dimensional black hole \cite{Witten, Gibbons:1992rh, Nappi, Creighton:1995uj, Davis:2004xi, Grumiller:2007ju, silence}.

Let us first look at the mass of the thermally excited solutions, (\ref{2dBH1}-\ref{2dBH2}), for the toy model of section \ref{toy}.  The ADM mass of this solution is (cf. Appendix \ref{ADM}) \be M_{ADM} = 8 e^a 4^8 V_8 \label{adm}.\ee
The factor of $4^8$ is due to our use of $y^i = \frac{x^i}{4}$ (compare (\ref{x1}) and (\ref{linear1})).
Here we assume that the $y^i$ coordinates range over a compact space of coordinate volume $V_8$. Another factor of 4 arises because our mass is defined by choosing $\xi = \frac{\partial }{\partial y^0}$ (as opposed to $\frac{\partial }{\partial x^0}$). With this understanding the result agrees with \cite{Witten, Gibbons:1992rh, Nappi, Creighton:1995uj, Davis:2004xi, Grumiller:2007ju, silence}.

We would like to verify that our counterterms construction also yields (\ref{adm}). To this end, we expand the solution (\ref{2dBH1}-\ref{2dBH2}) around the background (\ref{linear1}-\ref{linear2}). After a
short calculation we find the subleading terms in the expansion of the metric (in gauge (\ref{falloff1})):, \bea h^1_{tt} = \frac{6}{7} 4^8 e^a,  & &
h^1_{ii} = \frac{1}{7} 4^8 e^a, \\
h^1 = \frac{2}{7} 4^8 e^a,  & & \alpha = - \frac{2}{7} 4^8 e^a. \eea
To calculate expressions (\ref{toyT1}) and (\ref{toyjPhi1}) we also need the subleading terms in the expansions of $K_{ij}$ and $\hat K_{ij}$, \ie $p^1_{ij}$ and $\hat p^1_{ij}$. Expanding $K_{ij}$ yields
\be
p_{ij}^1 = -3 h^1_{ij}. \label{psimp}
\ee
The calculation of $\hat p^1_{ij}$ is slightly involved. We need to invert the relation (\ref{hatK:defn1}) perturbatively using the equations of motion.
The techniques to solve for $\hat p^1_{ij}$ were developed in \cite{MMV} (appendix B) in the context of  asymptotically flat spacetimes.  Performing a similar analysis for linear dilaton spacetimes and using the results of appendix \ref{sec:eom1order}, one finds
\be
\hat p^1_{ij} = h^1_{ij} - \frac{1}{4} h^0_{ij} h^1. \label{phatsimp}
\ee
A key step in this calculation is the use of equation (\ref{eom1order:toy}).

Using (\ref{psimp}) and (\ref{phatsimp}) we simplify the expressions (\ref{toyT1}) 
to get
\be
\tilde T_{ij} = \frac{1}{2} \left( - (9 h^1 + \al) h^0_{ij} + 16 h^1_{ij} \right). \label{toyT2}
\ee
Now we can easily calculate the mass and the other quantities from the counterterms. From the relation (\ref{Q2}) for the time translation Killing vector we find
\bea M_{ct} &=&
\int_C \sqrt{-h^0_C}  \tilde T_{ij} \xi^i n^j_C \\
&=& \int_C \sqrt{-h^0_C}  \tilde T_{tt} \xi^t n^t_C \\
&=& \int_C \left( -
\frac{1}{2}\left( 9 h^1 + \al\right) \ h^0_{tt}  + 8 h^1_{tt}
\right) \xi^t \left( \rho n^t \right) \\ &=& 8 e^a 4^8 V_8 =
M_{ADM},\eea
where in the third step we have used the expression (\ref{toyT2}).

One can similarly calculate the pressures (\ie the space-space components of $\tilde T_{ij}$). One finds the expected results \cite{Gibbons:1992rh, Nappi, Creighton:1995uj, Davis:2004xi, Grumiller:2007ju, silence},
\be
T_{ii} = 0 \label{pressure}.
\ee
We note that \cite{silence} arrived at (\ref{pressure}) by assuming the first law of thermodynamics
\be
dE = T dS.
\ee
We have therefore verified that this law does indeed holds for thermally excited linear dilaton solutions.

For the thermally excited solutions of the type II theory it is also possible to calculate the contribution of various terms in (\ref{stress}). It is a rather long calculation, but in the end it gives identical results \ie
\bea
M_{ct} &=& 8 e^a 4^5 V_5 (2 \pi^2), \\
T_{ii} &=& 0 \mbox{ (for directions along } \mathbb{R}^6 \mbox{)}.
\eea
and the first law $d M = T d S$ holds.

Thus, our counterterm formalism reproduces results previously known in the literature. For completeness, we have also calculated the scalar response function for both theories.  For the toy model one finds
\be
\tilde j_{\Phi} = \frac{1}{8} \left( - 9 h^1 + 47 \al \right) \label{toyjPhi2}
= - 2 e^a 4^8.
\ee
Again, one obtains the same final result in the type II theory via a somewhat longer calculation.

\section{Conclusions}
\label{conclusions}

Our work above addressed holographic renormalization of gravity in ten-dimensional asymptotically linear dilaton spacetimes dual to little string theory.  We have renormalized the action in the sense that i) the renormalized on-shell action is finite and ii) the renormalized action is fully stationary about solutions when the variations preserve asymptotically linear dilaton boundary conditions.    It would be interesting to understand in detail the relation between our counter-terms and those proposed in \cite{Davis:2004xi,Grumiller:2007ju} (based on a different construction) for two-dimensional linear dilaton spacetimes, in which the graviton itself has no local dynamics.

We have  computed the boundary stress tensor $T_{ij}$ and the scalar response function $j_\Phi$ for the toy model of section \ref{toy} and for the NS sector of type II theory.  The stress tensor is locally conserved, and gives the conserved charges which generate the asymptotic symmetries.  In particular, the rescaled limits $\tilde T_{ij} $ and $\tilde j_\Phi$ give finite quantities on the $\mathbb{R}^6$ which is the home of the LST.  We also showed that our expressions reproduce known results  \cite{Witten, Gibbons:1992rh, Nappi, Creighton:1995uj, Davis:2004xi, Grumiller:2007ju, silence} for the mass and pressure associated with thermally excited linear dilaton solutions.

However, these response functions are not quite the one-point functions of little string theory.  Note that each of $\tilde T_{ij} $ or $\tilde j_\Phi$ is built from the term in $T_{ij} $ and $j_\Phi$ associated with a {\it single} power of $\rho$.  In contrast, we expect that different fourier modes\footnote{Here we take $h^0_{ij}$ to be the flat metric $\eta_{ij}$.} $k$ of the one-point functions are associated with different powers of $\rho$; \ie with $\rho^\nu$ where $\nu = \nu(k^2)$.   Thus, we may expect that $\tilde T_{ij} $ and $\tilde j_\Phi$ give only certain Fourier modes of the one-point functions.  Since they integrate to the correct conserved quantities, one may guess that they correspond to the $k^2=0$ part of the one-point functions.  It would be interesting to understand this relation in detail, and to use our renormalized action to compute the full one-point functions.

After posting this work on the arxiv, we discovered that boundary terms for the closely related D5-brane spacetimes were also given in \cite{CO5}.  There the terms were given as part of a family of boundary terms for all Dp-brane spacetimes, and were constructed using techniques more familiar from the AdS context.  The results were used to compute the gravitational action ($I$) for Euclidean black 5-branes and the standard result $I=0$ was obtained.  Interestingly, the basic form of the boundary terms from \cite{CO5} appears quite different from ours, and is more reminiscent of AdS boundary terms.  While it is clearly of interest to understand to what extent our results are related to those of \cite{CO5}, here we simply note that $p=5$ is a special case for which many of the expressions in \cite{CO5} (e.g., the boundary stress tensor) diverge, and that no claim is made in \cite{CO5} regarding the staionarity of the action.  As a result, it is unclear whether a direct relation is expected.

Although holographic renormalization is most commonly studied for asymptotically AdS spacetimes, our work here was based on techniques developed for asymptotically flat spacetimes \cite{MM, MMV}. Their success in the linear dilaton context emphasizes the similarity between the two asymptotic behaviors.  This connection was recently used  to propose a framework for holographic duality in asymptotically flat spacetimes \cite{AF}, and further exploration of this link should provide additional insight in the future.

\subsubsection*{Acknowledgements}
We thank Keith Copsey for several useful discussions. This research was supported in part by the National Science Foundation under Grant No PHY03-54978, and by funds from the University of California.

\appendix
\setcounter{equation}{0}
\section{Boundary Stress Tensor Calculations}
\label{onepoint}

This section contains the details of various calculations needed to obtain the boundary stress tensor described in sections \ref{toy:renormalized} and \ref{NS:renormalized}.

\subsection{The Toy Model}
\label{onepoint:toy}
In this appendix we calculate
the expressions (\ref{toyT}) and (\ref{toyjPhi}) in detail. To this end, we find it convenient to separate the contributions of $S_0$ from those of the  counterterms. The contributions from $S_0$ are
\bea
\tilde T_{ij}^{orig} &=& -2 \left( p^1_{ij} - 9 h^1_{ij} - h^0_{ij} p^1 + h^0_{ij} h^1 \right),\label{Torig:toy}\\
\tilde j_\Phi^{orig} &=& 8 \al \label{jPhiorig:toy},
\eea
and the counterterm contributions are
\bea
\tilde T_{ij}^{ct} &=&  - 2 \left( 9 h^1_{ij} - h^1 h^{0}_{ij} + \hat p^1 h^0_{ij} - 2 \hat p^1_{ij} - 2 M^1_{ij} - h^1_{ij} - \frac{\al}{4} h^0_{ij}\right), \label{Tct:toy}\\
\tilde j_\Phi^{ct} &=& \frac{\al}{8} - 2 M^1_{\phi} \label{jPhict:toy},
\eea
where we have used the expansions
\bea
p_{ij} = \frac{1}{\rho} K_{ij} = h^0_{ij} + \frac{1}{\rho^\ell} p^1_{ij} +\frac{1}{\rho^{\ell+1}} p^2_{ij} + \ldots,  \ \ {\rm and} \label{p}\\
\hat p_{ij} = \frac{1}{\rho} \hat K_{ij} = h^0_{ij} + \frac{1}{\rho^\ell} \hat p^1_{ij} +\frac{1}{\rho^{\ell+1}} \hat p^2_{ij} + \ldots \label{hatp}\ \ .
\eea
In addition,  $M^1_{ij}$ and $M^1_{\phi}$ are coefficients in the expansion of $ h^{kl}\delta \hat K_{kl}$ defined through (see also equation (\ref{temp}))
\be
h^{kl} \delta \hat K_{kl} =  \ldots +  \frac{1}{\rho^{\ell-1}} M^1_{ij} \delta h^{ij} + \frac{1}{\rho^{\ell+1}} M^1_{\Phi} \delta \Phi  + \ldots \ \ .
\ee

Following \cite{MM} we  calculate the variation $h^{kl} \delta \hat K_{kl}$ as
a power series expansion of $\rho^{-1}$. Our starting point is the defining equation (\ref{hatK:defn1}) for $\hat K_{ij}$.  Taking the variation of (\ref{hatK:defn1}) we find
 \bea \delta \hat {\cal R}_{ij} &:=& \delta {\cal R}_{ij} +
\frac{e^{\Phi/2}}{2} \delta h_{ij} + \frac{e^{\Phi/2}}{4} h_{ij}
\delta \Phi  \\ &=& \delta \hat K_{kl} \left( h^{kl} \hat K_{ij} +
\delta_i^k \delta_j^l \hat K
- \delta_i^k  \hat K_j^l - \delta_j^k \hat K_i^l \right)
 + \left( \hat K_{ij} \hat K_{mn} - \hat K_{im} \hat K_{nj}
\right) \delta h^{mn} \label{invert:toy},\eea
which can be written in the form
\be \delta \hat {\cal R}_{ij} = L_{ij}{}^{kl} \delta \hat
K_{kl} + M_{ijkl} \delta h^{kl} \label{invert1}\ee  for \be
L_{ij}{}^{kl} = h^{kl} \hat K_{ij} + \delta_{i}^{k} \delta_j^l \hat
K - \delta_i^k \hat K_j^l - \delta_j^k \hat K_i^l, \ee and \be
M_{ijmn} = \hat K_{ij} \hat K_{mn} - \hat K_{im} \hat K_{nj}.\ee
We can invert the relation (\ref{invert1}) to get
\be
\label{main}
h^{kl}
\delta \hat K_{kl} = h^{mn} \left(L^{-1}\right)_{mn}{}^{ij}
\left[ \delta \hat {\cal R}_{ij} - M_{ijkl} \delta h^{kl} \right].\ee

We will evaluate (\ref{main}) by performing an expansion of various pieces in
$\frac{1}{\rho}$.  In particular, we introduce the expansions
\be L = L^0 + \frac{L^1}{\rho^{\ell}} + \dots  \mbox{ and }
\left(L^{-1}\right) = \left(L^{-1}\right)^0 + \frac{\left(L^{-1}\right)^1}{\rho^{\ell}} + \dots,
\ee with a similar expansion for $M$. The index 1 refers to the next-to-leading order term in the expansion.
A simple calculation shows that
\be \left(L^{-1}\right)^1{}_{ij}{}^{pq} = -
\left(L^{-1}\right)^0{}_{ij}{}^{kl} \left(L^1\right)_{kl}{}^{mn}
\left(L^{-1}\right)^0{}_{mn}{}^{pq} \ee
where,
\bea
\left(L^{-1}\right)^0{}_{ij}{}^{kl} &=& \frac{\rho}{7} \left[
\delta_i^k
\delta_j^l - \frac{1}{16} h^0{}^{kl}h^0_{ij}\right], \mbox{ and } \\
\left(L^1\right)_{ij}{}^{kl} &=& \frac{1}{\rho} \Bigg{[} \hat
p^1_{ij} h^0{}^{kl} - h^1{}^{kl} h^0_{ij} + \delta_i^k \delta_j^l
(\hat p^1 - h^1 ) \nonumber \\ & & + \left( h^0_{jm} h^1{}^{ml} - \hat
p^1_{jm} h^0{}^{ml} \right) \delta_i^k + \left( h^0_{im} h^1{}^{ml}
- \hat p^1_{im} h^0{}^{ml} \right) \delta_j^k\Bigg{]}.\eea The first
expression is given in \cite{MM} and the second expression
can be easily calculated by performing an expansion in $\rho^{-1}$.
With this information at hand we find
\be h^{kl} \delta \hat K_{kl} =
\frac{1}{\rho^2} \left( h^0{}^{mn} - \frac{h^1{}^{mn}}{\rho^\ell}
 + \ldots \right) \left[ (L^{-1})^0 + \frac{(L^{-1})^1}{\rho^\ell} +
 \ldots
\right]_{mn}^{ij} \left(\delta \hat {\cal R}_{ij} - M_{ijkl} \delta
h^{kl}\right). \label{BKL} \ee

Now, equation (\ref{BKL}) contains  two types of terms. One type comes with a factor of
$\delta \hat {\cal R}_{ij}$, and the other comes with a factor of $M_{ijkl} \delta h^{kl}$.
The leading-order terms of the first type are:
\be \frac{1}{\rho^2} h^0{}^{mn} (L^{-1})^0 _{mn}{}^{ij} \delta \hat
{\cal R}_{ij} - \frac{1}{\rho^{\ell+2}} h^1{}^{mn} (L^{-1})^0 _{mn}{}^{ij} \delta \hat
{\cal R}_{ij}+ \frac{1}{\rho^{\ell+2}} h^0{}^{mn} (L^{-1})^1 _{mn}{}^{ij} \delta \hat
{\cal R}_{ij}. \ee
After some calculation one may write these terms in the form
\be \frac{1}{16 \rho} h^0{}^{ij}\delta \hat
{\cal R}_{ij} - \frac{1}{7\rho^{\ell+1}} \left[ h^{1}{}^{ij} - \frac{h^1}{16} h^0{}^{ij} \right]\delta \hat
{\cal R}_{ij}- \frac{1}{112 \rho^{\ell+1}} \left[  -7h^{1}{}^{ij} - 2 \hat p^1{}^{ij} + \hat p ^1 h^0 {}^{ij} \right] \delta \hat {\cal R}_{ij}.\ee
Substituting the expansion for $\hat {\cal R}_{ij}$ and discarding the total derivative terms we find
\bea
h^{mn}\left(L^{-1}\right)_{mn}{}^{ij} \delta \hat {\cal R}_{ij} &=& - \frac{\rho}{2} h^0_{ij} \delta h^{ij} + \frac{9}{4 \rho} \delta \Phi + \frac{1}{28} \left[ \left( 2 \hat p^1 - 2 h^1 - 7 \al\right) h^0_{ij} - 10 h^1_{ij} - 4 \hat p^1_{ij} \right] \frac{\delta h^{ij}}{\rho^{\ell-1}} \nonumber \\ & +& \frac{1}{8} \left( 2 h^1 - 2 \hat p^1 + 9 \al\right) \frac{\delta \Phi}{\rho^{\ell+1}} + \ldots \ \ .
\eea

Let us now move on to the terms with $M_{ijkl} \delta h^{kl}$. We first
note the expansion for $M_{ijkl}$ yields \bea M^0_{ijkl} &=& \rho^{2}
\left( h^{0}_{ij} h^{0}_{kl} - h^{0}_{ik} h^{0}_{jl} \right), \\
M^1_{ijkl} &=& \rho^{2} \left( h^0_{ij} \hat p^1_{kl} +  h^{0}_{kl} \hat
p^1_{ij} - h^0_{ik} \hat p^1_{jl} -
h^0_{jl} \hat p^1_{ik} \right).\eea The term we need to expand is (cf.
(\ref{main})) \be - h^{mn} \left( L^{-1} \right)_{mn}{}^{ij} M_{ijkl}
\delta h^{kl}, \ee and the result produces  four terms:
\bea  & &-h^{mn} \left( L^{-1} \right)_{mn}{}^{pq} M_{pqkl}
\delta h^{kl}  =  -  \frac{h^{0}{}^{mn}}{\rho^2} (L^{-1})^0_{mn}{}^{pq} M^0_{pqkl} \delta
h^{kl}+ \frac{1}{\rho^{\ell+2}} h^{1}{}^{mn} (L^{-1})^0_{mn}{}^{pq}
M^0_{pqkl} \delta h^{kl} \nonumber \\ & & - \frac{1}{\rho^{\ell+2}} h^{0}{}^{mn} (L^{-1})^1_{mn}{}^{pq}
M^0_{pqkl} \delta h^{kl} - \frac{1}{\rho^{\ell+2}} h^{0}{}^{mn} (L^{-1})^0_{mn}{}^{pq}
M^1_{pqkl} \delta h^{kl} + \ldots \eea
After some calculation we find,
\bea -h^{mn} \left( L^{-1} \right)_{mn}{}^{pq} M_{pqkl}
\delta h^{kl}  &=& -  \frac{\rho}{2} h^{0}_{kl}  \delta
h^{kl} +  \frac{1}{7\rho^{\ell-1}} \left[ \frac{h^1}{2} h^0_{kl} -
h^1_{kl} \right]\delta h^{kl} \nonumber \\&+&
\frac{1}{112 \rho^{\ell-1} } \left[ - 7 h^1 h^0_{kl}
+ 6 \hat p^1 h^0_{kl} + 7 h^1_{kl} + 2 \hat p^1_{kl} \right] \delta h^{kl} \nonumber \\
&-& \frac{1}{16 \rho^{\ell-1}} \left[ \hat p^1 h^0_{kl} + 7
\hat p^1_{kl} \right] \delta h^{kl} + \ldots \\
=
- \frac{\rho}{2} h^0_{kl} \delta h^{kl} &+& \frac{1}{112} \left[ (h^1 - \hat p^1) h^0_{kl} - 9 h^1_{kl} - 47 \hat p^1_{kl} \right] \frac{\delta h^{kl}}{\rho^{\ell - 1}} + \ldots \ \ .
\eea
Combining the above results yields
\be \delta \hat K_{kl} h^{kl} = - \rho h^0_{ij} \delta h^{ij} + \frac{9}{4 \rho} \delta \Phi + M^1_{ij} \frac{\delta h^{ij}}{\rho^{\ell-1}} + M_{\Phi}^1 \frac{\delta \Phi}{\rho^{\ell+1}} + \ldots, \ee
with
\bea
M^1_{ij} &=& \frac{1}{16} \left[ ( -h^1 + \hat p^1 - 4 \al)h^0_{ij} - 7 h^1_{ij} - 9 \hat p^1_{ij} \right]\label{toy:m1h},\\
M^1_{\Phi} &=& \frac{1}{8} \left( 2h^1 - 2 \hat p^1 + 9 \al\right) \label{toy:m1phi}.
\eea
Substituting these expressions in (\ref{Tct:toy}) and (\ref{jPhict:toy}) and combining it with the contributions from the original action (\ref{Torig:toy}) and (\ref{jPhiorig:toy}) we get the desired result \ie equations (\ref{toyT1}) and (\ref{toyjPhi1}):
\bea
\tilde T_{ij} &=& \frac{1}{4} \left[ - (h^1 - 8 p^1 + 7 \hat p^1 + 2 \al)h^0_{ij} + h^1_{ij} - 8 p^1_{ij} + 7 \hat p^1_{ij} \right], \\
\tilde j_\Phi &=& \frac{1}{8} \left( 47 \al - 4 h^1 +  4 \hat p^1\right).
\eea


\subsection{NS Sector of Type II Theory}
\label{onepoint:NS}
We now perform the corresponding calculations for the full type II theory.  The general framework is the same as for the toy model above. The contribution from $S_0$  (equation (\ref{action})) to $\tilde j_{\Phi}$ is
\be
\tilde j_{\Phi}^{orig} = 8 \al,
\ee
and the counterterms contributions are
\bea
\tilde j_{\Phi}^{ct\:1} &=& - 2 M^1_{\Phi}, \\
\tilde j_{\Phi}^{ct\:2} &=& - \frac{75}{2} \al - \frac{15}{8} h^1 \cdot \Omega^0 + \frac{5}{8192} \beta^{1\:ijk}\beta^0_{ijk}, \\
\tilde j_{\Phi}^{ct\:3} &=&  - \frac{7}{4} h^1 \cdot \Omega^0 - \frac{7}{512}  {\cal R}^1 \cdot \Omega^0 + \frac{147}{4} \al - \frac{7}{16384} \beta^{1ijk}\beta^0_{ijk}, \mbox{ and}  \\
 \tilde j_{\Phi}^{ct\:4} &=& \frac{7}{8} \al,
\eea
where $M_1^\Phi$ is the contribution coming from $h^{kl} \delta \hat K_{kl}$ terms
\be
h^{kl}\delta \hat K_{kl} = \ldots  + \frac{1}{\rho^{\ell + 1}}M^1_{\Phi} + \ldots \ \ .
\ee
Using the techniques of the previous section one finds
\be
M^{\Phi}_1 = \frac{1}{56} \left[ 62 h^1 - 23 h^1\cdot \Omega^0 - 62 \hat p^1 + 2 \hat p^1 \cdot \Omega^0 - 273 \al\right] + \frac{1}{8192} \beta^{1\:ijk}\beta^0_{ijk} \label{M1Phi:NS},
\ee
where $h^1\cdot \Omega^0$ is the contraction of $h^{1ij}$ with the unit metric $\Omega^0_{ij}$ on $S^3$  \ie
\be h^1\cdot \Omega^0 = h^{1ij} \Omega^0_{ij}, \ee and similarly for $\hat p^1 \cdot \Omega^0$, etc. In obtaining the expression (\ref{M1Phi:NS}) we have used the expansion (\ref{hatp}) for $\hat K_{ij}$.
Combining the above results yields equation (\ref{jtildePhi:NS}),
\bea
\tilde j_{\Phi} = \frac{1}{56} (-124 h^1+39 h^1 \cdot \Omega^0 +124 \hat p^1 -4 (\hat p^1 \cdot \Omega^0 +\frac{49}{256} {\cal R} ^1 \cdot  \Omega^0 )+1001 \alpha)-\frac{\beta^{1 ijk}\beta^0_{ijk}}{16384}.
\eea

\setcounter{equation}{0}
\section{ADM Mass of the Thermally Excited Solution}
\label{ADM}

Following the Regge-Teitelboim construction \cite{RT} of the ADM surface terms \cite{ADM1} for the action (\ref{action1})
we obtain  \bea H[\xi] = \int d^8 S_d \sqrt{-q} G^{abcd} \left[ \xi^{\perp} \delta
q_{ab;c} - \xi^{\perp}_{,c} \delta q_{ab} \right] - \int d^8 S_d
\xi^{\perp} \sqrt{-q} D^d\Phi \delta \Phi, \eea
where
\bea
G^{abcd} = \frac{1}{2} \left( q^{ac} q^{bd} + q^{ad} q^{bc} - 2 q^{ab} q^{cd}\right),
\eea
and $q_{ab}$ is the induced metric on a Cauchy slice. Here the semicolon(;) denotes the torsion free covariant derivative with respect to the metric $q_{ab}$.  For the thermally excited solution (\ref{2dBH1}-\ref{2dBH2}) the scalar and the metric contributions to the ADM integrals for the timelike Killing field are (in gauge (\ref{falloff1}))
\bea \mbox{ scalar } &=& - \frac{8}{7} e^a 4^8 V_{8}, \\
\mbox{ metric } &=& \frac{64}{7} e^a 4^8 V_{8}, \eea which sum to
\be M_{ADM} = 8 e^a 4^8 V_{8} \label{mass1}. \ee
Similar calculations for the action (\ref{action}) and the solution (\ref{thermal1}-\ref{thermal3}) give
\be
M_{ADM} = 8 e^a V_5 4^5 (2 \pi^2)\label{mass2}.
\ee
In comparing (\ref{mass1}) and (\ref{mass2}), it is important to recall that, for the toy model, we rescaled all boundary coordinates to arrive at the metric (\ref{linear1}), but for the NS sector of the type II supergravity we only rescaled six of the coordinates, those corresponding to $\mathbb{R}^{6}$, to arrive at (\ref{bg1}). Thus, the factor $V_5 4^5 (2 \pi^2)$ in (\ref{mass2}) is in fact the natural analog of the factor $4^8 V_8$ in (\ref{mass1}).

\setcounter{equation}{0}
\section{Hypersurface Splitting of the Riemann Tensor}
\label{sec:eom1order}

In this section we review, for the sake of completeness, the Gauss-Codazzi splitting of the Riemann and the Ricci tensor. We use these results in section \ref{app:thermo}. Many of the results reviewed here are standard and can be found in, for example, \cite{FM}. Following the derivation of the Gauss-Codazzi equations \cite{Wald}, we note the following two relations.
When all indices on the bulk Riemann tensor $R_{\mu \nu  \rho \sigma}$ of $g_{\mu \nu}$ lie in the hypersurface one finds (Wald eq.~(10.2.23))
\be
R_{ijkl} = \: {\cal R}_{ijkl} - K_{ik} K_{jl} + K_{jk} K_{il}.
\ee
For all but one index in the hypersurface one finds
\bea
R_{ijk\perp} &=& h_{i}{}^{\mu} h_{j}{}^{\nu} h_{k}{}^{\s} \left( R_{\mu\nu \s \gamma} n^{\gamma} \right)= h_{i}{}^{\mu} h_{j}{}^{\nu} h_{k}{}^{\s} \left( \n_{\mu} \n_{\nu} n_{\s} - \n_{\nu} \n_{\mu} n_{\s} \right) \\
&=& D_i D_j n_k - D_j D_i n_k \\
&=& D_{i}K_{jk} - D_{j}K_{ik}
 \label{iden}
\eea
where $\perp$ denotes the direction perpendicular to the hypersurface (the radial direction $\rho$) and the derivative $D_i$ is the (torsion-free) covariant derivative on $\p M$ compatible with the boundary metric $h_{ij}$. Here we have used the Lemma 10.2.1 of \cite{Wald} in the second line and the definition of $K_{ij}$ in the last line. The identity (\ref{iden}) is a slight generalization of the second Gauss-Codazzi equations which are obtained by tracing over indices $i$ and $k$.

General expressions for the Riemann tensor with two indices perpendicular to the hypersurface for the metric (\ref{met}) are very complicated, though they simplify with our choice of Gaussian normal gauge ($N=1, N^i=0$).  In this gauge one may use direct calculation to find
\bea
\Gamma^{\perp}_{\perp \perp}  = 0,  \hspace{0.5 cm} \Gamma^\perp_{\perp i} = 0, & &  \Gamma^i_{jk} =^{(d-1)}\Gamma^i_{jk}, \\
\Gamma^{\perp}_{ij} = - K_{ij} &\mbox{ and }& \Gamma^i_{\perp j} = K^i{}_j .
\eea
Thus
\be
R_{i \perp j \perp} = - \frac{\partial K_{ij}}{\p \rho} + K_{ik}K^{k}{}_j .
\ee

Corresponding relations for the Ricci tensor follow immediately:
\bea
R_{ij} &=& {\cal R}_{ij} - K_{ij} K + 2 K_{ik}K^k{}_{j} - \frac{\p K_{ij}}{\p \rho}, \label{use}\\
R_{\perp i} &=& D_{j} K^j_i - D_i K, \\
R_{\perp \perp} &=& - h^{ij} \frac{\p K_{ij}}{\p \rho} + K_{ik}K^{k}{}_{j} h^{ij} \label{perp}.
\eea
Inserting (\ref{use}) in equation (\ref{toy:eom1}) and substituting the aysmptotic expansions for the metric, extrinsic curvature, and dilaton one obtains the first order dynamical equation of motion for the metric,
\be
\label{eom1order:toy}
{\cal R}^1_{ij} = - 4 ( \al + h^1 ) h^0_{ij}.
\ee

\end{document}